\documentclass[a4paper,fleqn]{cas-sc}

\usepackage[numbers]{natbib}

\def\tsc#1{\csdef{#1}{\textsc{\lowercase{#1}}\xspace}}
\tsc{WGM}
\tsc{QE}
\tsc{EP}
\tsc{PMS}
\tsc{BEC}
\tsc{DE}

\DeclareMathOperator{\argmax}{argmax} 
\DeclareMathOperator{\argmin}{argmin} 
\newcommand{\sign}{\text{sign}}

\usepackage{soul}
\soulregister\cite7
\soulregister\ref7
\soulregister\pageref7
\usepackage{xcolor}          
\usepackage[most]{tcolorbox} 

\definecolor{LightBlue}{rgb}{0.75,0.936,1.00}
\sethlcolor{LightBlue}
\definecolor{Yellow}{rgb}{1.00, 1.00, 0.2}
\sethlcolor{Yellow}
\definecolor{LightCyan}{rgb}{0.88,1,1}
\usepackage{float}
\usepackage{placeins}

\begin{document}

\shorttitle{ Revisiting TAS/GSC: A New Perspective and Application to DF Multihop Networks over Nakagami-$m$ Fading Channels}
\title{Revisiting TAS/GSC: A New Perspective and Application to DF Multihop Networks over Nakagami-$m$ Fading Channels}


\author[1]{Efendi Fidan}[type=editor,
                        ]
 \ead{efendifidan@gmail.com}

\credit{Conceptualization of this study, Methodology, Software}
\affiliation[1]{organization={TUBITAK National Metrology Institute (TUBITAK UME)},
                addressline={Gebze}, 
                city={Kocaeli},
                country={Turkey}}
                
\author[2]{Büşra Demirkol}[type=editor,
                        orcid=0000-0002-4851-9456]
 \ead{busrademirkol@gtu.edu.tr}

\credit{Data curation, Writing - Original draft preparation}

\affiliation[2]{organization={Department of Electronics Engineering, Gebze Technical University},
                addressline={Gebze}, 
                city={Kocaeli},
                country={Turkey}}

\author[2]{Oğuz Kucur}

\ead{okucur@gtu.edu.tr}

\credit{Data curation, Writing - Original draft preparation}





\begin{abstract}
In this work, transmit antenna selection (TAS) and generalized selection combining (GSC), i.e., TAS/GSC, are investigated over independent and identically distributed (i.i.d.) Nakagami-$m$ flat fading channels. An alternative analytical approach based on polynomial-power expansions and partial-fraction decomposition is developed to obtain finite-sum representations of the GSC and TAS/GSC output signal-to-noise ratio (SNR) distributions. These representations lead to closed-form expressions for outage probability (OP), symbol error rate (SER), and ergodic capacity. Compared with existing multinomial theorem-based formulations, the derived expressions provide a more structured representation of the coefficients and summation terms. As an example, the proposed approach is applied to decode-and-forward (DF) multihop networks with TAS/GSC at each hop over independent and non-identically distributed (i.n.i.d.) Nakagami-$m$ fading channels. Exact closed-form expressions for the end-to-end OP and ergodic capacity, together with an analytical lower bound on the average SER, are derived. The high-SNR outage behavior is also investigated to determine the diversity order. Finally, the accuracy of the derived analytical expressions is verified by Monte Carlo simulations. Numerical results show that significant performance improvements are obtained with TAS/GSC. Furthermore, the proposed analytical representation provides a basis for investigating related TAS/GSC-assisted multi-antenna relay configurations in emerging wireless networks.
\end{abstract}

\begin{keywords}
Transmit Antenna Selection (TAS), Receive Antenna Selection (RAS), Maximal-ratio Combining (MRC), Generalized Selection Combining (GSC), TAS/GSC, Decode-and-Forward, Multihop, Nakagami-$m$.
\end{keywords}
\maketitle

\section{Introduction}

Multihop relaying has been widely investigated to extend the coverage of wireless communication networks and alleviate the limitations imposed by path loss and shadowing \cite{hasna2003outage,boyer2004,soleimani2024multihop,khennoufa2022multihop}. In these networks, source information reaches the destination through intermediate relay nodes, providing an alternative to direct transmission when the source--destination link is weak, obstructed, or unavailable. Multihop relaying is relevant to wireless ad--hoc and mesh networks, connectivity in infrastructure-constrained areas, and wireless backhaul in cellular deployments \cite{boyer2004,lei2024mesh,polese2020,madapatha2020}. For example, integrated access and backhaul (IAB) in fifth-generation (5G) networks uses wireless links to connect access nodes without requiring a wired backhaul connection at every site \cite{polese2020,madapatha2020}. Nevertheless, the end-to-end performance of a multihop connection depends on the propagation conditions and reliability of the individual hops \cite{hasna2003outage,boyer2004,soleimani2024multihop}.

Among the commonly employed relaying protocols, amplify-and-forward (AF) relays amplify and forward the received signal, whereas decode-and-forward (DF) relays decode and re-encode the received information before forwarding it to the next hop \cite{boyer2004, soleimani2024multihop}. In serial DF multihop networks, end-to-end outage occurs when at least one hop fails to support the required transmission rate, while decoding errors at intermediate relays also affect the overall error performance \cite{boyer2004,soleimani2024multihop,kim2014relay,khennoufa2022multihop}. Accordingly, the performance of DF multihop networks has been investigated in terms of outage probability (OP), symbol error rate (SER), and ergodic capacity under various fading conditions. In \cite{kim2014relay}, the SER of DF multihop systems is characterized using symbol transition matrices for independent and non-identically distributed (i.n.i.d.) fading links. In \cite{yilmaz2015multihop}, analytical expressions for OP, SER, and ergodic capacity are derived for DF multihop networks over Nakagami-$m$ fading channels, considering different transmit and receive diversity schemes. In \cite{ghofrani2020aeu}, the outage and diversity performance of DF multihop networks with signals received from two preceding nodes is investigated over Nakagami-$m$ fading channels. More recently, \cite{subhashini2026df} analyzes the outage and average bit error performance of DF multihop Internet of Things (IoT) systems under generalized fading conditions in millimeter-wave (mmWave) environments. These studies provide a foundation for investigating diversity techniques that improve the reliability of individual hops and the overall multihop transmission.

To improve the reliability of individual hops, multiple-input multiple-output (MIMO) techniques exploit spatial diversity provided by multiple transmit and receive antennas \cite{molisch2004,yilmaz2015multihop}. However, processing signals from all available antenna branches in conventional multi-antenna architectures may require multiple radio-frequency (RF) chains, increasing hardware cost, power consumption, and implementation complexity \cite{molisch2004,sanayei2004}. Antenna selection offers a lower-complexity alternative by using a subset of the available antennas while retaining the benefits of spatial diversity \cite{sanayei2004}. Among the commonly employed schemes, transmit antenna selection (TAS) activates a single transmit antenna according to the channel conditions, whereas selection combining (SC) or receive antenna selection (RAS) chooses the receive branch with the highest instantaneous signal-to-noise ratio (SNR). In contrast, maximal-ratio combining (MRC) combines the signals received through all available branches to maximize the output SNR \cite{sanayei2004,yilmaz2015multihop}. Since SC requires fewer active RF chains than MRC but generally provides a lower combining gain, more flexible approaches that balance diversity performance and receiver complexity are of interest.

Generalized selection combining (GSC) provides such flexibility by selecting the $L$ receive branches with the highest instantaneous SNRs among $N$ available branches and combining them using MRC \cite{ma2004efficient}. Accordingly, SC and MRC are special cases of GSC for $L=1$ and $L=N$, respectively. By adjusting the number of combined branches, GSC offers a trade-off between diversity performance and receiver complexity \cite{ma2004efficient,cai2004}. This flexibility can be extended to the transmit side by combining GSC with TAS. In the resulting TAS/GSC scheme, the transmit antenna is selected to maximize the output SNR obtained through GSC at the receiver \cite{cai2004}. Consequently, TAS/GSC encompasses TAS/SC (or TAS/RAS, or equivalently joint transmit and receive antenna selection (JTRAS)) and TAS/MRC as special cases, enabling the exploitation of both transmit and receive diversity without simultaneously processing all available antenna branches. These characteristics make TAS/GSC a flexible approach for multi-antenna relay networks, where the performance of individual hops affects the reliability of end-to-end transmission.

Several studies have investigated antenna selection and diversity combining in multihop and relay-assisted MIMO networks. In \cite{lee2009}, the OP of an AF multihop system employing TAS/MRC is analyzed over Rayleigh fading channels, considering ideal relay gains. The study in \cite{abdelnabi2016} further examines TAS/MRC in AF multihop networks under co-channel interference and feedback delay. For DF relaying, \cite{toka2018df} investigates the outage performance of antenna selection schemes in dual-hop full-duplex MIMO networks over Nakagami-$m$ fading channels, whereas \cite{tu2022performance} analyzes TAS/SC, TAS/MRC, and MRT/MRC in multihop MIMO networks under short-packet communications.
More recently, \cite{quang2026tas} investigates the OP and throughput of TAS/SC-assisted DF multihop networks employing power-domain non-orthogonal multiple access (NOMA) over Rayleigh fading channels. In \cite{saikia2026df}, TAS/MRC and JTRAS are examined in dual-hop DF multi-relay systems over Fisher--Snedecor $F$ fading channels. TAS is also employed in \cite{dinh2025aeu} for an in-band full-duplex relay system with rate-splitting multiple access over Nakagami-$m$ fading channels, where ergodic capacity and energy efficiency are analyzed.

Beyond SC- and MRC-based schemes, TAS/GSC has also been investigated in different communication scenarios. In \cite{wang2014twc}, the probability density function (PDF) and cumulative distribution function (CDF) of the TAS/GSC output SNR are derived to evaluate the average secrecy rate and secrecy outage probability (SOP) of MIMO wiretap channels over Nakagami-$m$ fading. In \cite{deng2015tsp}, exact analytical expressions for OP, SER, and ergodic capacity are obtained for cognitive dual-hop DF relay networks employing TAS/GSC over Nakagami-$m$ fading channels. In \cite{lee2024hybrid}, a dual-hop AF-MIMO relay system employing TAS at the source and relay, GSC at the relay receiver, and SC at the legitimate user and eavesdropper is investigated over Rayleigh fading channels in terms of SOP. More recently, \cite{lee2025gsc} analyzes the secrecy performance of dual-hop DF relay systems over Nakagami-$m$ fading channels, considering GSC at both the relay and user receivers, TAS at the source and relay, and MRC at the eavesdropper. These studies demonstrate the applicability of antenna selection and GSC to different relay configurations and performance evaluation problems.

Beyond conventional relay configurations, multihop communication and multi-antenna techniques have also been investigated in emerging wireless networks. In particular, a multi-user multihop massive relay MIMO system is considered to improve coverage and resilience against blockages in mmWave communications in \cite{ke2025wcnc}. In \cite{an2026noma}, SC-based single-input multiple-output (SIMO) DF multihop networks incorporating NOMA, energy harvesting, and fountain coding are analyzed in terms of OP and average channel capacity. These studies highlight the continuing relevance of multihop relaying and antenna diversity techniques to emerging communication systems.

Although the statistical properties of GSC and TAS/GSC over Nakagami-$m$ fading channels have been characterized in previous studies, their exact analytical treatment involves nontrivial manipulations of ordered SNR statistics \cite{annamalai2003,wang2014twc,deng2015tsp}. In particular, the TAS/GSC formulations in \cite{wang2014twc,deng2015tsp} employ multinomial expansions involving multiple summation indices and parameter-dependent coefficients, which can complicate their application to more general network configurations. This complexity becomes particularly relevant in serial DF multihop networks, where the statistical expressions of individual hops must be combined to evaluate end-to-end performance. An earlier conference study \cite{fidan2021performanceMultihop} investigates the asymptotic outage performance of serial DF multihop networks employing TAS/GSC over Nakagami-$m$ fading channels. However, that study focuses on asymptotic OP rather than a broader analysis involving exact OP, error performance, and ergodic capacity. These observations motivate an alternative analytical representation of GSC and TAS/GSC statistics that facilitates their application to multihop networks. Specifically, a structured finite-sum representation with systematically calculated polynomial coefficients is desirable for analyzing serial DF multihop systems with i.n.i.d. fading conditions across hops.

Motivated by these observations, this paper investigates a serial half-duplex DF multihop network employing TAS/GSC at each hop over Nakagami-$m$ fading channels. The channel gains are assumed to be independent and identically distributed (i.i.d.) within each hop, while the fading parameters may vary across hops. We develop an alternative finite-sum representation of the GSC and TAS/GSC output-SNR statistics and apply it to the end-to-end performance analysis of DF multihop networks. The main contributions of this work are summarized as follows:

\begin{enumerate}
    \item An alternative MGF-based derivation is developed for GSC over i.i.d. Nakagami-$m$ fading channels with positive integer-valued shape parameters. Using polynomial-power expansions and partial-fraction decomposition, the output-SNR CDF is expressed as a finite sum of exponential terms with polynomial coefficients, from which closed-form expressions for OP, average SER, and ergodic capacity are obtained.

    \item The resulting GSC representation is extended to TAS/GSC through sequential binomial and polynomial-power expansions. Closed-form expressions for the TAS/GSC output-SNR CDF, OP, average SER, and ergodic capacity are obtained, including the single-branch selection case.

    \item By applying TAS/GSC at each hop of a serial DF multihop network with i.n.i.d. fading conditions across hops, exact expressions for the end-to-end OP and ergodic capacity are derived. An analytical lower bound on the end-to-end average SER of $M$-ary phase shift keying ($M$-PSK) and square $M$-ary quadrature amplitude modulation ($M$-QAM) modulations is also obtained, which becomes tight in the high-SNR region.

    \item The high-SNR outage behavior is characterized to determine the end-to-end diversity order. The derived analytical expressions are validated through Monte Carlo simulations, and the effects of fading severity and antenna selection configurations on system performance are examined.

    \item By developing alternative representations of TAS/GSC statistics and applying to DF multihop networks, this work provides more tractable analysis of TAS/GSC implementations in recent and future systems \cite{ke2025wcnc, an2026noma}.
\end{enumerate}

The remainder of this paper is organized as follows. In Section \ref{ChapVISystemModel}, details of system model are provided. Section \ref{PerformanceSection}  includes closed-form derivations. Numerical results are elaborated in  Section \ref{ChapVINumericalResults}. Finally, we conclude our work in Section \ref{ChapVIConclusion}.

\section{System Model} \label{ChapVISystemModel} 
\begin{figure}[pos=t]
	\centering
	\includegraphics[width=5in]{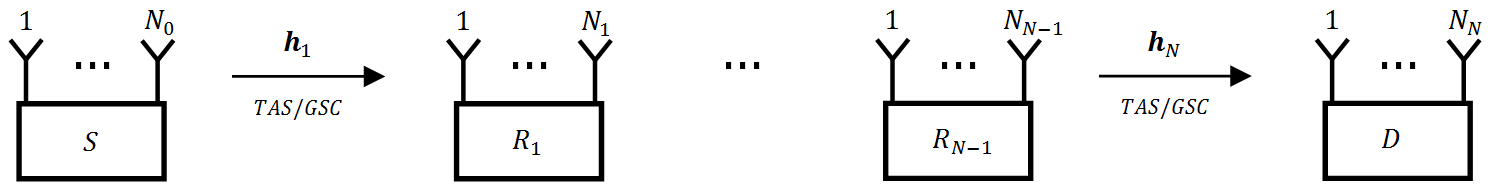}
	\caption{Multihop system with TAS/GSC.}
	\label{SystemFigure}
\end{figure} 

 Performance of a multi antenna DF multihop system consisting of $N+1$ nodes ($N$ hops) employing TAS/GSC in each hop is investigated over Nakagami-$m$ flat fading channels, which is demonstrated in Fig. \ref{SystemFigure}. The first node is a multi-antenna source ($S$, $1^{st}$ node),  the last node is a multi-antenna end-user/destination ($D$, $\left(N+1\right)^{th}$ node), and  $N-1$  intermediate nodes are half-duplex multi-antenna relays ($R_i, i \in\{1,\dots, N-1\}$) employing DF transmission protocol. Data transmission from $S$ to $D$ is accomplished in orthogonal bands (relays receive and transmit data in orthogonal bands) or two phases:  In the first phase, $S$ and even numbered relays broadcast while other relays are in silence. In the second phase odd numbered relays  broadcast while  $S$ and even numbered relays are in silence. Total number of antennas at each node is represented by $N_i$, $i \in\{0,\dots, N\}$. So, $N_0$ is the total number of antennas at $S$. The transmitted signal of $S$ and estimated signals at the relays are assumed to have average unit  energy and line-of-sight (LOS) is only available in between successive nodes. In hop-$i$, the channel gain between $k^{th}$ transmit antenna and $j^{th}$ receive antenna is $h_{j,k}^i \text{ } \bigl(i\in\{1\dots, N\}, \text{ } k\in\{1\dots, N_{i-1}\} \text{ and } j\in\{1\dots, N_{i}\}\bigr)$. The channels between the successive nodes are assumed to be i.i.d. Nakagami-$m$ distributed and the ones among hops are modelled as  i.n.i.d. Nakagami-$m$ distributed. Hence, shape and scale factors at each hop are represented by $m_i$ and $\Omega_i$, $i \in\{1,\dots, N\}$, respectively. To improve readability, descriptions of the parameters are given in Table \ref{ParametersDescriptions}.

\begin{table}[!t]
\scriptsize
\centering
\caption{Parameters / Functions Descriptions.}
\begin{tabular}{|l|l|}
\hline
\textbf{Parameters / Functions} & \textbf{Descriptions} \\
\hline
$N_i$ &
\begin{tabular}{@{}l@{}}Total number of antennas at each node, $i \in\{0,\dots, N\}$\end{tabular} \\
\hline
$L_i$
 & Total number of selected antennas at $i^{th}$ hop \\
\hline
$h_{j,k}^i$ & The channel gain between $k^{th}$ transmit antenna and $j^{th}$ receive antenna at $i^{th}$ hop   \\
\hline
$m_i$,  $\Omega_i$ & Shape and scale factors at $i^{th}$ hop \\
\hline
$\gamma_{j,k}^i$ & The instantaneous SNR between the $k^{th}$ transmit and $j^{th}$ receive antennas  at $i^{th}$ hop \\
\hline
$P_i$ & Total transmitted power  at $i^{th}$ hop \\
\hline
$\sigma_i^2$  & Variance of additive Gaussian noise  at $i^{th}$ hop \\
\hline
$\lambda_i$  & Inverse of the expected value of $\gamma_{j,k}^i$ \\
\hline
$\beta_{k}^{i}$  &  SNR of GSC at $i^{th}$ hop \\
\hline
$\gamma_i$ & SNR of TAS/GSC at $i^{th}$ hop  \\
\hline
$\textbf{h}_{i}$ & Channel gain vector  of TAS/GSC  at $i^{th}$ hop
\\
\hline
 $ f_{\gamma_{j,k}^i}(x)$ &  PDF of $\gamma_{j,k}^i$
\\
\hline
 $ F_{\gamma_{j,k}^i}(x)$ & CDF of $\gamma_{j,k}^i$ \\
\hline
$\Phi_{\beta_{k}^{i}}(s)$  & MGF of $\beta_{k}^{i}$  \\
\hline
$\Phi_{\gamma_{j,k}^i}(s,x)$  & Complementary MGF of $\gamma_{j,k}^i$ \\
\hline
$\Gamma(\alpha)$  & Gamma function \\
\hline
$\gamma(\alpha,x)$  & Lower incomplete Gamma function \\
\hline
$\Gamma(\alpha,x)$  & Upper incomplete Gamma function \\
\hline
$\mu_i^{\Phi}(q, L_i-1, m_i-1)$  & 
Coefficients in the polynomial expansion of $[\Phi_{\gamma_{j,k}^i}(s,x)]^{L_i-1}$ \\
\hline
$\mu_i^{F}\left(j, n, m_i-1\right)$ & 
Coefficients in the polynomial expansion of $[F_{\gamma_{j,k}^i}(x)]^{N_i-L_i}$\\
\hline
\end{tabular}
\label{ParametersDescriptions}
\end{table}

Transmission is conducted via simultaneous selection of a single transmit antenna and  $L_i$ receive antenna(s) at each hop, namely, TAS/GSC is implemented. The instantaneous SNR between the $k^{th}$ transmit and $j^{th}$ receive antennas, the total transmitted power, and variance of additive Gaussian noise (AGN) at hop $i$ are $\gamma_{j,k}^i=P_i |h_{j,k}^i|^2/\sigma_i^2$, $P_i$, and    $\sigma_i^2$, respectively. In that case, the expected values of all instantaneous SNRs at each hop and their inverses are $P_i \Omega_i/\sigma_i^2$ and  $\lambda_i=\sigma_i^2/(P_i \Omega_i)$, respectively. Generalized antenna selection is carried out at each relay and $D$ based on instantaneous SNRs in two stage: In the first stage, these SNRs,  $\{\gamma_{j,k}^i\}_{j=1}^{N_{i}}$, are sorted in descending order to obtain order statistics $\gamma_{k,1:N_i}^i\geq\gamma_{k,2:N_i}^i\geq\cdots\geq\gamma_{k,N_i:N_i}^i$. Thereafter, first $L_i$  of these order statistics are added up to obtain $\beta_{k}^{i}=\sum_{l_i=1}^{L_i}\gamma_{k,l_i:N_i}^i$, which is the SNR of GSC. In the second stage, the maximum sum SNR (SNR of TAS/GSC) is determined as $\gamma_i=\max\{\beta_{k}^{i}\}_{k=1}^{N_{i-1}}$ and the index of  transmit antenna producing this maximum SNR is attained via  $k_i=\argmax\{\beta_{k}^{i}\}_{k=1}^{N_{i-1}}$, which is fed back to the preceding node to initiate the transmission. We assume the error-free feedback and the perfect channel information at the receiver part. In the receiving node, incoming signals of the selected $L_i$  antennas are merged via MRC and the estimated signal is broadcast to the next node in the corresponding transmission phase.

\section{Performance Analysis} \label{PerformanceSection} 

 Closed-form expressions of OP, SER, and ergodic capacity for GSC, TAS/GSC, and multihop TAS/GSC are provided in this section. 

Two important identities that used many times during the derivations are binomial expansion and power expansion of a finite polynomial. Hence, it is beneficial to give the identity for power expansion of a finite polynomial \cite[eq. (0.314)]{jeffrey2007table}:
  \begin{equation} \label{Denklem_1}
\left(\sum_{q=0}^{m}\left(a_q x^j\right) \right)^{p}=\sum_{q=0}^{m p}\left(\mu_x\left(q, p, m\right) x^q\right)  
  \end{equation}
where $\mu\left(0, p,m\right)=a_0^p$, $\mu\left(q, p,m\right)=\frac{1}{q a_0}\sum_{k=1}^{q}[(kp-q+k)a_k\mu\left(q-k, p,m\right)]$, and $a_k=0$ for $k>m$. Additionally, illustrating the methodology how the binomial expansion is implemented throughout this work eases understanding of our derivations. As an example assume binomial expansion is applied to the term $(x+y+z)^p$. In the first step,  the term $x+y+z$ is partitioned into two groups, namely, $x$ and $y+z$, thereafter binomial expansion is implemented to get $(x+y+z)^p=\sum_{k_0=0}^{p}\binom{p}{k_0} x^{p-k_0}(y+z)^{k_0}$. In the second step, binomial expansion is applied to the term $(y+z)^{k_0}=\sum_{k_1=0}^{k_0}\binom{k_0}{k_1} y^{k_0-k_1}z^{k_1}$, in turn,  $(x+y+z)^p=\sum_{k_0=0}^{p}\sum_{k_1=0}^{k_0}\binom{p}{k_0} \binom{k_0}{k_1} x^{p-k_0} y^{k_0-k_1}z^{k_1}$.

The channel gains are assumed to be Nakagami-$m$ distributed, hence, PDF and CDF of $\gamma_{j,k}^i$ (SNR of each link in the $i^{th}$ hop)  are 
\begin{equation} \label{Denklem_2}
\begin{aligned}
f_{\gamma_{j,k}^i}(x)=&\frac{\left(m_i\lambda_i\right)^{m_i}}{\Gamma(m_i)}x^{m_i-1}e^{-m_i\lambda_ix}
\end{aligned}
\end{equation}
and
\begin{equation} \label{Denklem_3}
\begin{aligned}
F_{\gamma_{j,k}^i}(x)=
\frac{\gamma\left(m_i,m_i\lambda_ix\right)}{\Gamma(m_i)}=
1-e^{-m_i\lambda_ix} \sum_{j=0}^{m_i-1} \left( \frac{(m_i\lambda_ix)^j}{\Gamma(j+1)} \right),
\end{aligned}
\end{equation}
respectively, where Gamma function $\Gamma(\alpha)=\int_0^\infty t^{\alpha-1}e^{-t}dt$ \cite[eq. (8.310.1)]{jeffrey2007table} and lower incomplete Gamma function $\gamma(\alpha,x)=\int_{0}^{x}t^{\alpha-1}e^{-t}dt$ \cite[eq. (8.350.1)]{jeffrey2007table}. The equality in the second line of (\ref{Denklem_3}) is valid for integer values of $m_i$ \cite[eq. (8.352.1)]{jeffrey2007table}.

In the remaining parts of this section, firstly, OP, SER, and ergodic capacity expressions for GSC are provided. Secondly, expressions for TAS/GSC are given and finally, multihop TAS/GSC expressions are introduced.

\subsection{OP, SER, and Ergodic Capacity of GSC}

CDF of $\beta_{k}^{i}$ (SNR of GSC) is derived based on MGF method \cite[eq. (5)]{ma2004efficient}.  To get closed-form of CDF of $\beta_{k}^{i}$, $F_{\beta_{k}^{i}}(x)$, firstly the identity given in (\ref{Denklem_1}) is applied to $[\Phi_{\gamma_{j,k}^i}(s,x)]^{L_i-1}$ and  $[F_{\gamma_{j,k}^i}(x)]^{N_i-L_i}$ (its implementation to this term is done after a binomial expansion), where $\Phi_{\gamma_{j,k}^i}(s,x)$ is the complementary MGF of $\gamma_{j,k}^i$ \cite[eq. (3)]{ma2004efficient} (refer to (\ref{Denklem_A5}) and (\ref{Denklem_A5_1})). Secondly closed-form MGF of $\beta_{k}^{i}$, $\Phi_{\beta_{k}^{i}}(s)$, is obtained. In sequence, PDF of  $\beta_{k}^{i}$, $f_{\beta_{k}^{i}}(x)$, is derived as in (\ref{Denklem_A9}) by taking inverse Laplace transform of $\Phi_{\beta_{k}^{i}}(s)$ which is converted to a simple version with implementation of partial fraction decomposition. Hence, CDF of $\beta_{k}^{i}$ becomes as in (\ref{Denklem_A10}). Finally, using the identity \cite[eq. (8.352.1)]{jeffrey2007table} and the fact that constant terms sum up to $1$, $F_{\beta_{k}^{i}}(x)$ is rearranged as in (\ref{Denklem_4}) (Please refer to Appendix A for more details).
  \begin{equation} \label{Denklem_4}
  \begin{aligned}
 F_{\beta_{k}^{i}}(x)=1-\sum_{n=0}^{N_i-L_i}\biggl[ P_{in}(x)e^{-\lambda_{in}x} \biggr]=
 1-\sum_{n=0}^{N_i-L_i}\biggl[ \sum_{j=0}^{T_{in} }  \left(c_{nj}^i x^j\right) e^{-\lambda_{in}x} \biggr].
   \end{aligned}
  \end{equation}
The polynomials, $P_{in}(x)$ ($P_{i0}(x)=\sum_{j=0}^{m_iL_i-1}\left(c_{0j}^i x^j\right)$ and $P_{in}(x)=\sum_{j=0}^{(m_i-1)(L_i+n)}\left(c_{nj}^i x^j\right)$), in (\ref{Denklem_4}) are  given in (\ref{Denklem_5}) and (\ref{Denklem_6}) for $n=0$ and $n\geq 1$, respectively, where $c_{nj}^i$ represents the coefficients of rearranged polynomials. This representation means CDF of GSC is a weighted sum of exponential functions with polynomial coefficients. The maximum degrees (the largest exponents) of the coefficient polynomials of $e^{-\lambda_{i0}x}$ and $e^{-\lambda_{in}x}$ are $m_iL_i-1$ and  $(m_i-1)(L_i+n)$, respectively. Therefore, the upper bounds of the inner summation in (\ref{Denklem_4}) are  $T_{in}=\left(m_iL_i-1\right)u[-n]+(m_i-1)(L_i+n)u[n-1]$, and $\lambda_{in}=\frac{(L_i+n)m_i\lambda_i}{L_i}$, where $u[n]$ is unit step function. Note that the calculation of coefficients $c_{nj}^i$ is straightforward, which can be easily implemented in any programming language.
  \begin{equation} \label{Denklem_5}
  \begin{aligned}
 P_{i0}(x)=&\sum_{q=0}^{(m_i-1)(L_i-1)}\sum_{k=0}^{m_i L_i-1} \left(C_{0,\beta_{k}^{i}}^{\Phi} \frac{\lambda_{i0}^{-m_i L_i+k} }{\Gamma(k+1)} x^k \right)+
\sum_{\lambda_{i0}} \left(C_{\lambda_{i0}}C_{\beta_{k}^{i}}^{\Phi} \frac{\lambda_{i0}^{-p+k} }{\Gamma(k+1)} x^k \right),
  \end{aligned}
  \end{equation}
where the shorthand notation 
$\sum_{\lambda_{i0}}=\sum_{q=0}^{(m_i-1)(L_i-1)}  \sum_{n=1}^{N_i-L_i} \sum_{j=0}^{n(m_i-1)} \sum_{p=1}^{m_i(L_i-1)-q} \sum_{k=0}^{p-1}$ and 
  \begin{equation} \label{Denklem_6}
  \begin{aligned}
P_{in}(x)=&\sum_{q=0}^{(m_i-1)(L_i-1)}   \sum_{j=0}^{n(m_i-1)} \sum_{p=1}^{q+j+m_i} \sum_{k=0}^{p-1} \left(C_{\lambda_{in}}C_{\beta_{k}^{i}}^{\Phi}   \frac{\lambda_{in}^{-p+k} }{\Gamma(k+1)} x^k  \right).
  \end{aligned}
  \end{equation}
The coefficients emerging in (\ref{Denklem_5}) and (\ref{Denklem_6}) are provided in (\ref{Denklem_7}), (\ref{Denklem_8}), (\ref{Denklem_9}), and (\ref{Denklem_10}).
  \begin{equation} \label{Denklem_7}
  \begin{aligned}
C_{\beta_{k}^{i}}^{\Phi}=\frac{(-1)^n(m_i\lambda_i)^{L_im_i} \Gamma(q+j+m_i)}{\Gamma(m_i) L_i^{q+j+m_i-1} }\binom{N_i}{L_i}\binom{N_i-L_i}{n}\mu_i^{\Phi}(q, L_i-1, m_i-1)  \mu_i^{F}\left(j, n, m_i-1\right),
  \end{aligned}
  \end{equation}
where $\mu_i^{\Phi}(q, L_i-1, m_i-1)$ are the coefficients due to polynomial expansion of $[\Phi_{\gamma_{j,k}^i}(s,x)]^{L_i-1}$ and $\mu_i^{F}\left(j, n, m_i-1\right)$ are the coefficients due to polynomial expansion of $[F_{\gamma_{j,k}^i}(x)]^{N_i-L_i}$.
  \begin{equation} \label{Denklem_8}
  \begin{aligned}
C_{0,\beta_{k}^{i}}^{\Phi}=\frac{(m_i\lambda_i)^{L_im_i} \Gamma(q+m_i)}{\Gamma(m_i) L_i^{q+m_i-1} }\binom{N_i}{L_i}\mu_i^{\Phi}(q, L_i-1, m_i-1)
  \end{aligned}
  \end{equation}
The coefficients related to the partial fraction decomposition in the derivation of the simplified MGF, $\Phi_{\beta_{k}^{i}}(s)$, are represented by $C_{\lambda_{in}}$  and are given in (\ref{Denklem_9}) and (\ref{Denklem_10}) for the poles $\lambda_{i0}$ ($n=0$) and $\lambda_{in}$ ($n\geq 1$), respectively. 
  \begin{equation} \label{Denklem_9}
  \begin{aligned}
  C_{\lambda_{i0}}=&\frac{(-1)^{m_i(L_i-1)-q-p}\Gamma(j+m_iL_i-p)  }{\Gamma(q+j+m_i)\Gamma\left(m_i(L_i-1)-q-p+1\right)}\left(\frac{nm_i\lambda_i}{L_i}\right)^{-\left(j+m_iL_i-p\right)}
  \end{aligned}
  \end{equation}
  and
  \begin{equation} \label{Denklem_10}
  \begin{aligned}
  C_{\lambda_{in}}=&\frac{(-1)^{m_i(L_i-1)-q}\Gamma(j+m_iL_i-p)  }{\Gamma(m_i(L_i-1)-q)\Gamma\left(q+j+m_i-p+1\right)}\left(\frac{nm_i\lambda_i}{L_i}\right)^{-\left(j+m_iL_i-p\right)}.
  \end{aligned}
  \end{equation}
This completes the derivation of the GSC outage expression.

Ergodic capacity of GSC is derived based on CDF method \cite[eq. (45)]{yang2013performance}. Using  the identity given in \cite[eq. (39)]{fidan2017performance}, the closed-form expression of ergodic capacity turns to (refer to (\ref{Denklem_A14}) and \ref{Denklem_A15}))
  \begin{equation} \label{Denklem_11}
  \begin{aligned}
 R_{GSC}=& \frac{\log_2(e)}{2}\sum_{n=0}^{N_i-L_i}\sum_{j=0}^{T_{in} }  \left(c_{nj}^i e^{\lambda_{in}} \Gamma(j+1)\Gamma(-j,\lambda_{in})\right), 
   \end{aligned}
  \end{equation}
where $\Gamma(\alpha,x)=\int_{x}^{\infty}t^{\alpha-1}e^{-t}dt$ is upper incomplete Gamma function \cite[eq. (8.350.2)]{jeffrey2007table}.

Average SER of $M$-PSK and $M$-QAM can be evaluated based on MGF method by \\ $\sum_{q=1}^{Q} a_q \int_{0}^{\theta_q} \Phi_{\beta_{k}^{i}}\left(\frac{\lambda_{\rm mod} }{\sin^2(\theta)}\right)d\theta$ \cite[eq. (5.3)]{simon2005digital}. The closed-form expression of SER is derived as (refer to (\ref{Denklem_A17}) and (\ref{Denklem_A18}))
 \begin{equation} \label{Denklem_12}
  \begin{aligned}
 P_{GSC}=\sum_{q=1}^{Q}\left(a_q\theta_q\right)-\sum_{q=1}^{Q} \sum_{n=0}^{N_i-L_i} \sum_{j=0}^{T_{in} }  \left(a_q c_{nj}^i \lambda_{\rm mod} \Gamma(j+1) 
 I_{j}(\theta_q; \lambda_{\rm mod}; \lambda_{in})\right),
   \end{aligned}
  \end{equation}
where the modulation type dependent parameters are given in Table \ref{SERTable}. The integral in (\ref{Denklem_12}) is defined and solved by aid of identity \cite[{eq. (38)}]{fidan2017performance} as
  \begin{equation} \label{Denklem_13}
  \begin{aligned}
 I_{m}(\theta_q; \lambda_{\rm mod}; \lambda_{in})=&\int_{0}^{\theta_q} \frac{\left(\sin^2(\theta)\right)^m}{ (\lambda_{\rm mod}+\lambda_{in}\sin^2(\theta))^{m+1}}d\theta\\=&
 \frac{ \dfrac{\sqrt{\pi}\Gamma[m+1/2]}{2\Gamma[m+1]} -\cos(T)_2F_1(\dfrac{1}{2},\dfrac{1}{2}-m;\dfrac{3}{2};\cos^2(T))}{(\lambda_{\rm mod})^{1/2}(\lambda_{\rm mod}+\lambda_{in})^{m+1/2}},
   \end{aligned}
  \end{equation}
where $T=\frac{1}{2}\tan^{-1}{\left(\frac{N}{D}\right)}+\frac{1}{\pi}\left(1-\sign(N)\frac{1+\sign(D)}{2}\right)$, $N=2\sqrt{\lambda_{\rm mod}(\lambda_{\rm mod}+\lambda_{in})}\sin(2\theta_q)$,  $D=(2\lambda_{\rm mod}+\lambda_{in})\cos(2\theta_q)-\lambda_{in}$ \cite[{eq. (5A.31) and (5A.32)}]{simon2005digital}. This integral is valid for $m>-1/2$ and $\sin(\varphi)\geq 0$ and $_2F_1(\iota,\tau;w;z)$ is the Gauss hypergeometric function \cite[{eq. (9.111)}]{jeffrey2007table}.  For integer values of $m$, another closed-form is also available \cite[eq. (2.513.1)]{jeffrey2007table}. This completes the analysis of GSC.

 \begin{table}  [!t]
\caption{Modulation type dependent parameters for  $M$-QAM and $M$-PSK SER calculations.}
\centering
\begin{tabular}{|c|c|c|c|}
\hline
{Parameter}  & {$M$-QAM} & {$M$-PSK} \\ \hline
{$a_1$}  & {$\frac{\pi}{4}\left(1-\frac{1}{\sqrt{M}}\right)$} & {$ \frac{1}{\pi} $}   \\ \hline
{$a_2$}    & {$-\frac{\pi}{4} \left(1-\frac{1}{\sqrt{M}}\right)^2$} & {0}         \\ \hline
{$\theta_1$}  & {$\frac{\pi}{2}$}  & {$\frac{(M-1)\pi}{M}$}     \\ \hline
{$\theta_2$}   & {$\frac{\pi}{4}$}               & {0}             \\ \hline
{$\lambda_{mod}$}         & {$\frac{3}{2\left(M-1\right)}$}          & {$\sin^2\left(\pi/M\right)$} \\ \hline
{$Q$}                          & {2}                   & {1}       \\ \hline
\end{tabular}
\label{SERTable}
\end{table}

\subsection{OP, SER, and Ergodic Capacity of TAS/GSC}

Using the derived GSC CDF in hand means closed-form derivation of TAS/GSC CDF,  $F_{\gamma_i}(x)=F_{\beta_{k}^{i}}(x)^{N_{i-1}}$, can be accomplished in two steps without using multinomial theorem but instead using the notion implemented in \cite[eq. (12)]{fidan2016performance}, where a new closed-form of TAS/GSC CDF over Rayleigh fading channels is introduced: In the first step, product of each exponential function and its polynomial coefficient, namely,  $P_{in}(x)e^{-\lambda_{in}x}$ in (\ref{Denklem_4}) is treated as a single term   and, then,  $N_i-L_i+1$ binomial expansions are carried out sequentially. Thereafter, in the second step, the identity given in (\ref{Denklem_1}) is applied to the resulting powers of polynomials, $P_{in}(x)$, which gives the closed-form of $F_{\gamma_i}(\gamma_{th})$. Applying $N_i-L_i+1$ binomial expansion to $F_{\gamma_i}(x)=F_{\beta_{k}^{i}}(x)^{N_{i-1}}$ in sequence results in 
  \begin{equation} \label{Denklem_14}
  \begin{aligned}
 F_{\gamma_{i}}(x)=\sum_{t_{i0}=0}^{N_{i-1}} \sum_{t_{i1}=0}^{t_{i0}} \cdots \sum_{t_{i{(N_i-L_i)}}=0}^{t_{i{(N_i-L_i-1)}}}\biggl[ (-1)^{t_{i0}} \Psi_{t_{in}} e^{-\Lambda_ix}  \prod_{n=0}^{N_i-L_i-1}[P_{in}(x)]^{t_{in}-t_{i(n+1)}}[P_{i{(N_i-L_i)}}(x)] ^{t_{i{(N_i-L_i)}}} \biggr],
   \end{aligned}
  \end{equation}
where $\Lambda_i=\frac{m_i\lambda_i\left(L_i t_{i0}+\sum_{n=1}^{N_i-L_i}t_{in} \right)}{L_i}$ and  $\Psi_{t_{in}}=\binom{N_{i-1}}{t_{i0}} \binom{t_{i0}}{t_{i1}} \cdots \binom{t_{i(N_i-L_i-1)}}{t_{i(N_i-L_i)}}$. Applying polynomial expansion given in (\ref{Denklem_1}) to the powers of polynomials emerging in (\ref{Denklem_14}) converts $F_{\gamma_{i}}(x)$ to
  \begin{equation} \label{Denklem_15}
  \begin{aligned}
 F_{\gamma_{i}}(x)=&1-\sum_{t_{in}}\sum_{k_{in}}\biggl[ (-1)^{t_{i0}+1} \Psi_{t_{in}} \mu_{i{(N_i-L_i)}}\left(k_{i{(N_i-L_i)}}, t_{i{(N_i-L_i)}}, T_{i(N_i-L_i)}\right)  \\& \times  \prod_{n=0}^{N_i-L_i-1}\left[\mu_{in}\left(k_{in}, t_{in}-t_{i(n+1)}, T_{in}\right)\right]  x^{K_i} e^{-\Lambda_ix} \biggr],
   \end{aligned}
  \end{equation}
where $\sum_{t_{in}}=\sum_{t_{i0}=1}^{N_{i-1}} \sum_{t_{i1}=0}^{t_{i0}} \cdots \sum_{t_{i{(N_i-L_i)}}=0}^{t_{i{(N_i-L_i-1)}}}$,  $K_i=\sum_{n=0}^{N_i-L_i}k_{in}$, and  $a_k^{P_{in}(x)}=c_{nj}^iu[T_{in}-k]$, 
$\sum_{k_{in}} = \sum_{k_{i0}=0}^{[t_{i0}-t_{i1}]T_{i0} }  \sum_{k_{i1}=0}^{[t_{i1}-t_{i2}]T_{i1}} \\ \cdots \sum_{k_{i{(N_i-L_i-1)}}=0}^{[t_{i{(N_i-L_i-1)}}-t_{i{(N_i-L_i)}}]T_{i{(N_i-L_i-1)}}} \sum_{k_{i{(N_i-L_i)}}=0}^{t_{i(N_i-L_i)}T_{i(N_i-L_i)}}$. Note that due to the last term of binomial expansion, there is no difference in the upper limit of last summation of $\sum_{k_{in}}$. Another essential point to mention is that the first term of $F_{\gamma_{i}}(x)$ corresponding to the case $t_{i0}=0$, where all $t_{in}$ become zero, is 1. This condition is explicitly written to facilitate derivations of ergodic capacity and CDF of multihop TAS/GSC. Thereby, $t_{i0}$ in the first summation of $\sum_{t_{in}}$ starts from 1. Furthermore, by defining a new coefficient, $F_{\gamma_{i}}(x)$ simplifies   as
  \begin{equation} \label{Denklem_16}
  \begin{aligned}
 F_{\gamma_{i}}(x)=1-&\sum_{t_{in}}\sum_{k_{in}}\bigl[ C_{\gamma_{i}} x^{K_i} e^{-\Lambda_ix} \bigr],
   \end{aligned}
  \end{equation}
where the coefficient $ C_{\gamma_{i}}$ is given as
  \begin{equation} \label{Denklem_17}
  \begin{split}
 C_{\gamma_{i}}=(-1)^{t_{i0}+1} \Psi_{t_{in}} \mu_{i{(N_i-L_i)}}\left(k_{i{(N_i-L_i)}}, t_{i{(N_i-L_i)}}, T_{i(N_i-L_i)}\right) \prod_{n=0}^{N_i-L_i-1}\left[\mu_{in}\left(k_{in}, t_{in}-t_{i(n+1)}, T_{in}\right)\right].
   \end{split}
  \end{equation}
Note that CDF of TAS/GSC has the same form as that of GSC. This means derivations of ergodic capacity and average SER are straightforward. Since it is trivial, there is no need to give MGF of TAS/GSC. Instead SER and ergodic capacity expressions are directly attained by imitating those of GSC.  Hence, closed-form ergodic capacity of TAS/GSC turns to
  \begin{equation} \label{Denklem_18}
  \begin{aligned}
R_{TAS/GSC}=&\frac{\log_2(e)}{2} \sum_{t_{in}}\sum_{k_{in}}\bigl( C_{\gamma_{i}} e^{\Lambda_i} \Gamma[K_i+1] \Gamma[-K_i, \Lambda_i]\bigr).
   \end{aligned}
  \end{equation}
Similary, SER expression of TAS/GSC becomes
  \begin{equation} \label{Denklem_19}
  \begin{aligned}
 P_{TAS/GSC}=& \sum_{q=1}^{Q}\left(a_q\theta_q\right)-\sum_{q=1}^{Q}\sum_{t_{in}}\sum_{k_{in}}\bigl[ a_q \lambda_{\rm mod} C_{\gamma_{i}} \Gamma(K_i+1) I_{K_i}\left(\theta_q;\lambda_{\rm mod};\Lambda_{i} \right)   \bigr].
   \end{aligned}
  \end{equation}
To complete the analysis of GSC and TAS/GSC, a last step is remained: The provided exact expressions do not work in case of single receive antenna selection, i.e., for $L_i=1$ but they are readily available in the literature. For completeness of analyses in this work, we provide CDF expression separately, in turn, ergodic capacity and SER expressions are straightforward.  Since $F_{\gamma_{i}}^{L_i=1}(x)=[F_{\gamma_{j,k}^i}(x)]^{N_i N_{i-1}}$, applying a polynomial expansion after a binomial expansion produces its closed-form given  as
  \begin{equation} \label{Denklem_20}
  \begin{aligned}
 F_{\gamma_{i}}^{L_i=1}(x)=&1-\sum_{t_{i0}=1}^{N_i N_{i-1}} \sum_{k_{i0}=0}^{t_{i0}(m_i-1)} \left(\binom{N_i N_{i-1}}{t_{i0}} (-1)^{t_{i0}+1} \mu_i^{F}\left(k_{i0}, t_{i0}, m_i-1\right)  x^{k_{i0}} e^{-t_{i0}m_i\lambda_ix} \right).
   \end{aligned}
  \end{equation}
Note that this representation coincides with that of (\ref{Denklem_16}). To avoid overloading of equations, we define $C_{\gamma_i}^{L_i=1}=\binom{N_i N_{i-1}}{t_{i0}} (-1)^{t_{i0}+1} \mu_i^{F}\left(k_{i0}, t_{i0}, m_i-1\right)$, which means ergodic capacity and SER expressions are as those of TAS/GSC given in (\ref{Denklem_18}) and (\ref{Denklem_19}), respectively. Of course, coefficients  $C_{\gamma_{i}}$, $K_i$, and $\Lambda_i$ should be replaced with $C_{\gamma_i}^{L_i=1}$,  $k_{i0}$, and $t_{i0}m_i\lambda_i$ , respectively, together with summation terms. For GSC, only $N_{i-1}=1$ assignment is needed. This ends up analyses of GSC and TAS/GSC.

\subsection{Multihop TAS/GSC Analysis}

Due to the implementation of DF transmission protocol, overall system outage is equivalent to the outage of a single hop in between. Therefore, the outage scheme is  corresponding to $\gamma_{eq}=\min\{\gamma_1, \gamma_2,\cdots,\gamma_N\}< \gamma_{th}$, where  $\gamma_{th}=2^{2R}-1$  is the SNR threshold for data rate $R$. So OP of multihop TAS/GSC can be evaluated as
\begin{equation} \label{Denklem_21}
  \begin{aligned}
F_{\gamma_{eq}}(\gamma_{th})&=Pr(\min\{\gamma_1, \gamma_2,\cdots,\gamma_N\}<\gamma_{th})\\
&=1-\prod_{i=1}^{N}Pr(\gamma_i>\gamma_{th})\\
&=1-\prod_{i=1}^{N}\left(1-F_{\gamma_i}(\gamma_{th}) \right).
  \end{aligned}
  \end{equation}
To get closed-form of $F_{\gamma_{eq}}(x)$, TAS/GSC CDFs given in (\ref{Denklem_16}) and (\ref{Denklem_20}) for each hop are inserted into (\ref{Denklem_21}) to get

\begin{equation} \label{Denklem_22}
\begin{aligned}
F_{\gamma_{eq}}(x)=&1-\prod_{i=1}^{N}\left(\sum_{t_{in}}\sum_{k_{in}}[ C_{\gamma_{i}} x^{K_i} e^{-\Lambda_ix}]\right)\\
=&1-\sum_{t_{1n}}\sum_{k_{1n}}\cdots\sum_{t_{(N_i-L_i)n}}\sum_{k_{(N_i-L_i)n}}\biggl(
\prod_{i=1}^{N}\left( C_{\gamma_{i}}\right) x^{\sum_{i=1}^{N} K_i} e^{-\sum_{i=1}^{N}\Lambda_ix}\biggr).
\end{aligned}
\end{equation}

For clarity and simplicity, we define $C_{\Lambda}=\prod_{i=1}^{N}\left( C_{\gamma_{i}}\right)$, $K=\sum_{i=1}^{N} K_i$, $\Lambda=\sum_{i=1}^{N}\Lambda_i$, and the shorthand $\sum_{\Lambda}=\sum_{t_{1n}}\sum_{k_{1n}}\cdots\sum_{t_{(N_i-L_i)n}}\sum_{k_{(N_i-L_i)n}}$. Thereafter, $F_{\gamma_{eq}}(x)$ converts to
\begin{equation} \label{Denklem_23}
\begin{aligned}
F_{\gamma_{eq}}(x)=&1-\sum_{\Lambda}\biggl(
C_{\Lambda} x^{K} e^{-\Lambda x}\biggr).
\end{aligned}
\end{equation}
This pretty simple representation actually means  derivations of ergodic capacity and SER are also completed, which are like those of GSC and TAS/GSC:
  \begin{equation} \label{Denklem_24}
  \begin{aligned}
R_{eq}=&\frac{\log_2(e)}{2} \sum_{\Lambda}\bigl( C_{\Lambda}  e^{\Lambda} \Gamma[K+1] \Gamma[-K, \Lambda]\bigr)
   \end{aligned}
  \end{equation}
  and
  
  \begin{equation} \label{Denklem_25}
  \begin{aligned}
 P_{eq}=& \sum_{q=1}^{Q}\left(a_q\theta_q\right)-\sum_{q=1}^{Q}\sum_{\Lambda} \bigl[ a_q \lambda_{\rm mod} C_{\Lambda} \Gamma(K+1) I_{K}\left(\theta_q;\lambda_{\rm mod};\Lambda \right)   \bigr].
   \end{aligned}
  \end{equation}
The preceding results provide the OP, ergodic capacity, and SER expressions for the considered systems. However, a critical point that should be clarified more about the SER expression given in (\ref{Denklem_25}) is actually a lower bound of SER of DF multihop system, i.e., it is assumed the link between the source and destination is equivalent to the weakest link. To be more precise, assuming $\gamma_{eq}=\min\{\gamma_1, \gamma_2,\cdots,\gamma_N\}$  and representing the channel gains of the weakest link/hop for TAS/GSC by $\textbf{h}_{I_{eq}}$ where $I_{eq}=\argmin\{\gamma_i\}_{i=1}^{N}$
means overall link  between the source and destination is $\textbf{h}_{I_{eq}}$. However, especially in lower SNR region, this does not take place in practice. Hence, the provided expression is a lower bound of exact SER expression. The equality becomes valid in the high SNR region. For more details please refer to \cite{belbase2019coverage}.

Lastly, denoting maximum CDF of $N$ hops by $F_{\gamma_i}^{m}(x)=\max\{F_{\gamma_i}(x)\}_{i=1}^{N}$ and representing their equality case as  $F_{\gamma}(x)=F_{\gamma_1}(x)=\cdots=F_{\gamma_N}(x)$ allows one to obtain lower and upper bounds of OP, $F_{\gamma_{eq}}(\gamma_{th})$, at high SNR region as \cite[eq. (12)]{fidan2021performanceMultihop}
\begin{equation} \label{Denklem_26}
  \begin{aligned}
      F_{\gamma_i}^{m}(\gamma_{th}) \leq F_{\gamma_{eq}}(\gamma_{th}) \leq N \times F_{\gamma}(\gamma_{th}),
  \end{aligned}
\end{equation}
which follows from the fact that product of two or more CDFs ($F_{\gamma_i}(\gamma_{th})$) can be omitted, since they are less than 1. Taking this fact into the consideration,  in case of no diversity scheme the slope of $log-log$ plot is $\min\{m_i\}_{i=1}^{N}$, where the index of the hop gives this slope is represented by $I_{e}=\argmin\{m_i\}_{i=1}^{N}$. After implementation of diversity scheme, the slope will be $\min\{m_iN_iN_{i-1}\}_{i=1}^{N}$ and the hop index is denoted by  $I_{y}=\argmin\{m_iN_iN_{i-1}\}_{i=1}^{N}$.  Therefore, the diversity order is  $\frac{m_{I_{y}}}{m_{I_{e}}}N_{I_{y}}N_{{I_{y}}-1}$ \cite{fidan2021performance}.

\section{Numerical Results} \label{ChapVINumericalResults}


In this section, numerical results for OP, SER, and ergodic capacity are provided to validate the derived expressions. Unless otherwise specified, equal transmit powers are allocated to the $N$ hops, such that $P_T=N P_i$, and the noise variances are normalized to $\sigma^2=\sigma_{R_i}^2=\sigma_D^2=1$. Depending on the considered scenario, the results are plotted versus the first-hop SNR, $P_0/\sigma^2$, or the total transmit SNR, $P_T/\sigma^2$, as indicated in each figure. The average channel powers are normalized to $\Omega_i=1$ for all hops, unless otherwise specified. All parameter settings are provided in the explanation of each figure. The analytical results are obtained from the corresponding GSC, TAS/GSC, and multihop expressions derived in Section~\ref{PerformanceSection}. In particular, (\ref{Denklem_23}), (\ref{Denklem_24}), and (\ref{Denklem_25}) provide the end-to-end OP, ergodic capacity, and SER lower bound of the multihop network, respectively.

\begin{figure}[pos=!t]
    \centering
    \includegraphics[width=5in]{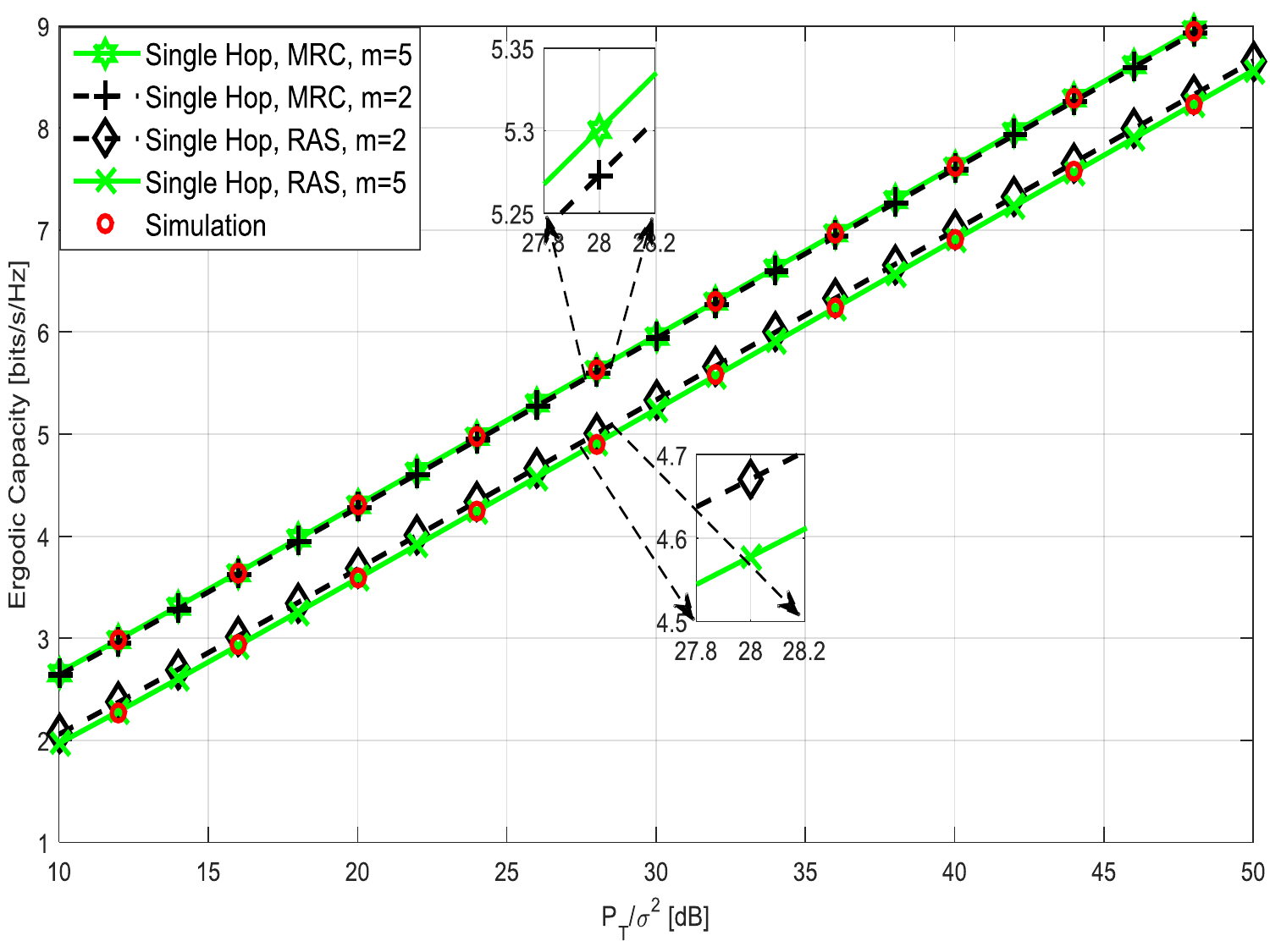}
    \caption{Ergodic capacity of single-hop GSC for special cases of RAS and MRC.}
    \label{GSCErgodicCapacity}
\end{figure}

Fig. \ref{GSCErgodicCapacity} illustrates the ergodic capacity of single-hop GSC with only two extreme antenna selection cases, namely, RAS and MRC for clarity. Total number of transmit and receive antennas are 1 and 4, respectively. Two shaping factors, $m \in \{2,5\}$, are illustrated. The capacity differences between the two fading parameters are relatively small, as highlighted in the magnified regions. For the sake of clarity, the simulation results are solely provided for $m=5$ and in excellent agreement with the analytical results. Interestingly, increasing the Nakagami-$m$ shape parameter slightly improves the ergodic capacity of MRC, whereas the opposite behavior is observed for RAS. For MRC, increasing $m$ reduces the variance of the combined SNR while preserving its mean, resulting in a slight improvement in ergodic capacity. In contrast, increasing $m$ reduces the variations among the individual branch SNRs, thereby decreasing the average selection gain achieved by RAS. Consequently, a slight reduction in ergodic capacity is observed.

\begin{figure}[pos=!t]
    \centering
    \includegraphics[width=5in]{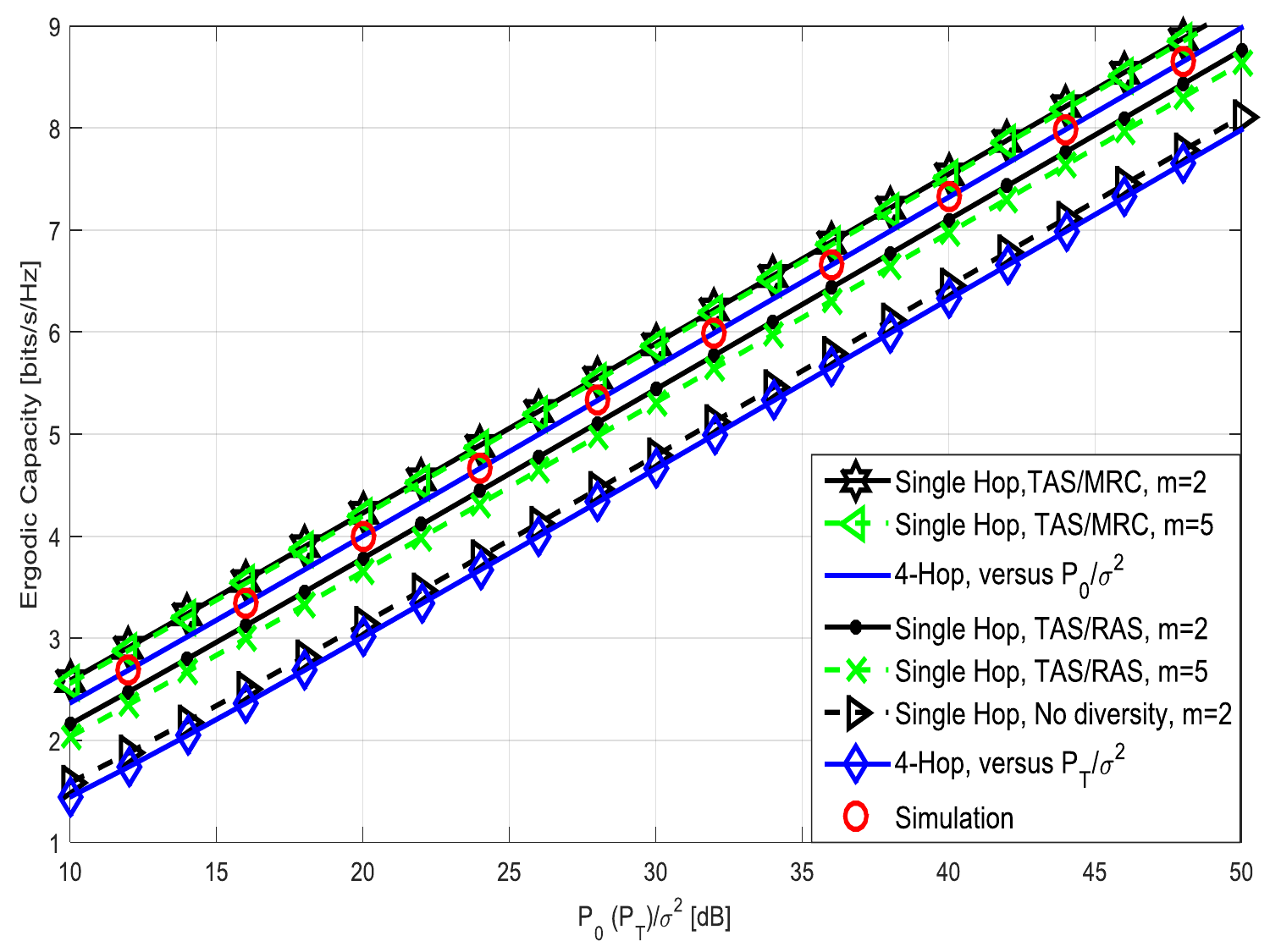}
    \caption{TAS/GSC ergodic capacity for single-hop and four-hop transmissions.}
    \label{TAS-GSCErgodicCapacity}
\end{figure}

In Fig. \ref{TAS-GSCErgodicCapacity}, the ergodic capacity of the TAS/GSC scheme is illustrated for single-hop and 4-hop transmissions. Parameter settings for single-hop TAS/GSC are $m \in \{2,5\}$, $N_0=2$, $N_1=3$, and $L_i\in \{1, 3\}$, i.e., TAS/RAS and TAS/MRC are implemented. Additionally, a single hop with no diversity scheme for $m=2$ is also provided for comparison. In contrast to the sole GSC/MRC case in Fig. \ref{GSCErgodicCapacity}, increasing the shape factor from $m=2$ to $m=5$ slightly decreases the ergodic capacity of both TAS/RAS and TAS/MRC. This behavior is mainly attributed to the reduction in TAS gain as the fading variations decrease. Comparison of TAS/MRC and the case with no diversity scheme for $m=2$ reveals that an increment of approximately 1.1 bits/Hz/s in capacity is attained. For 4-hop transmission, parameters are $(m_1, m_2, m_3, m_4)=(1, 2, 3, 2)$, $(N_0, N_1, N_2, N_3, N_4)=(3, 3, 3, 3, 4)$, $(L_1, L_2, L_3, L_4)=(2, 2, 2, 4)$, which means TAS/MRC is conducted in the fourth hop and TAS/GSC is done in the other hops. For the sake of clarity, simulation results are only provided for 4-hop transmission and their overlapping with analytical results is excellent. As expected, the 4-hop capacity curve plotted versus the total SNR, $P_T/\sigma^2$, is shifted to the right compared with that plotted versus the first-hop SNR, $P_0/\sigma^2$, due to the total transmit power being distributed among four hops. Furthermore, when the total SNR is considered, at the price of larger coverage the 4-hop system exhibits a lower ergodic capacity than the single-hop system without diversity, despite employing TAS/GSC at each hop. This result is mainly due to the distribution of the total transmit power among the hops and the dependence of the end-to-end performance on the weakest hop.

\begin{figure}[pos=!t]
    \centering
    \includegraphics[width=5in]{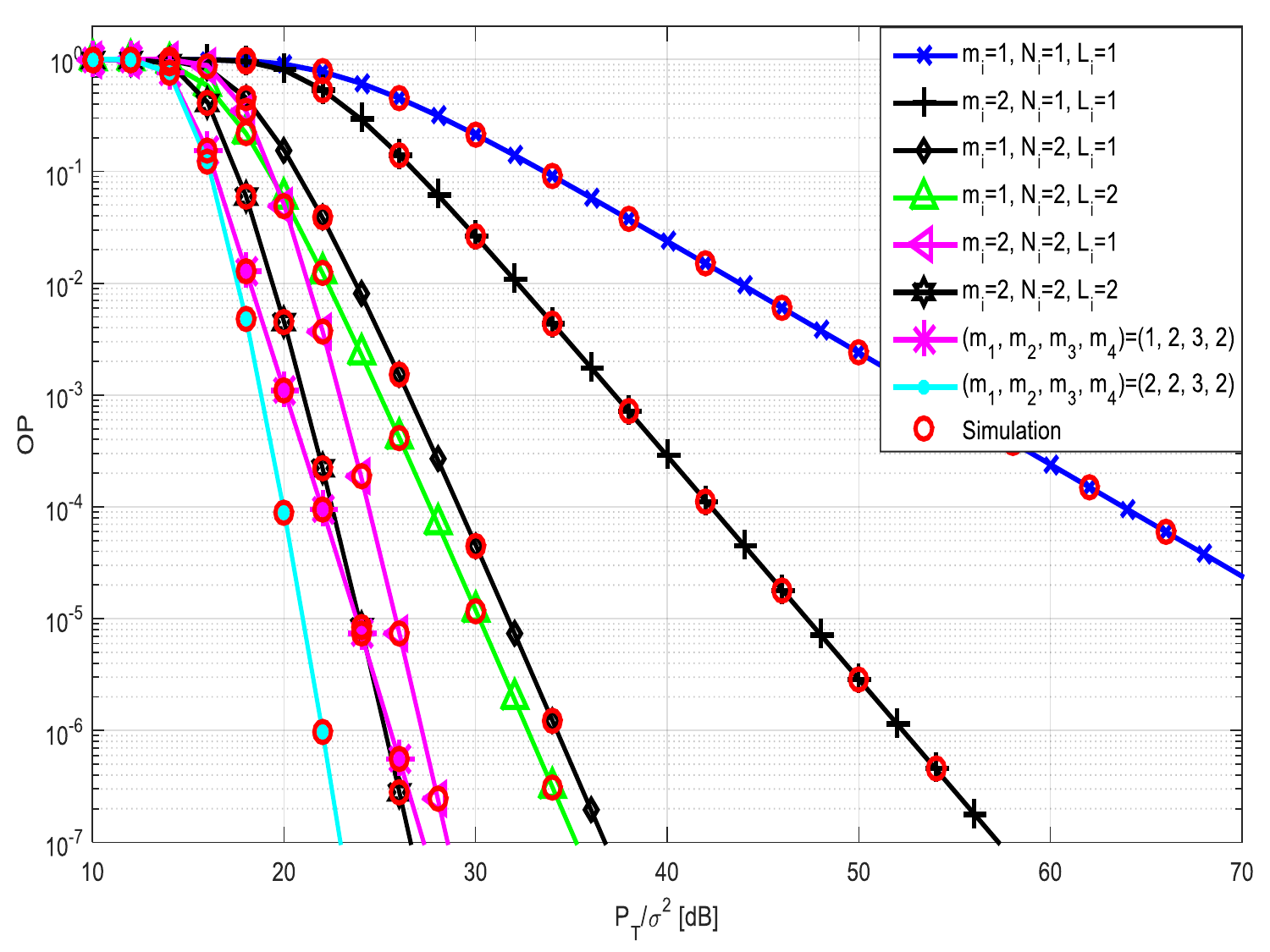}
    \caption{Multihop OP for different Nakagami-$m$ parameters and antenna configurations.}
    \label{MultiHopOP}
\end{figure}

Fig. \ref{MultiHopOP} verifies the accuracy of the OP expression given in (\ref{Denklem_23}). The analytical and simulation results are in excellent agreement. Total number of hops is 4 and the target data rate is set to 2 bits/s/Hz. The two upper curves correspond to the cases without any diversity schemes, i.e, $N_i = L_i=1$, for Rayleigh fading ($m_i=1$, $i \in \{1, 2, 3, 4\}$) and Nakagami-$m$ fading ($m_i=2$). 
In the second scenario, the total number of transmit and receive antennas is set to $N_i=2$ for $m_i \in \{1, 2\}$, where TAS/RAS or equivalently JTRAS (number of selected receive antennas $L_i=1$, $i \in \{1, 2, 3, 4\}$) (black lines with diamond markers for $m_i=1$ and magenta lines with left-pointing triangle markers for $m_i=2$) and TAS/MRC ($L_i=2$) (green lines with upward triangle markers for $m_i=1$ and black lines with star markers for $m_i=2$) antenna selection schemes are illustrated. Compared with the case without diversity, TAS/RAS provides an SNR gain of approximately 35 dB for Rayleigh fading ($m_i=1$) at an OP value of $10^{-4}$. The corresponding gain for $m_i=2$ is about 18 dB. Comparison of TAS/MRC and TAS/RAS schemes of that scenario for shape factors $m_i=1$ and $m_i=2$ at an OP value of $10^{-5}$ shows that the gain in total SNR is about 1.8 dB. This improvement is achieved by combining the signals from both receive antennas instead of selecting only the strongest one. In the third scenario, schemes with different shape factors and total number of transmit and receive antennas with $L_i=2$ are demonstrated, where $(m_1, m_2, m_3, m_4) \in \{(1, 2, 3, 2), (2, 2, 3, 2)\}$ (magenta lines with asterisk markers and cyan lines with circle markers, respectively), $N_0=N_2=N_4=2$, and $N_1=N_3=3$. TAS/GSC in the first and third hops and TAS/MRC in the second and fourth hops are implemented. Comparison of curves with different shape factors at an OP value of $10^{-6}$ reveals that increasing the shape factor of the first hop ($m_1$) from 1 to 2 results in a gain of about 4.3 dB in total SNR. This improvement is expected since the weakest link is the first hop with shape factor $m_1=1$. Increasing $m_1$ improves the reliability of this hop and increases the end-to-end diversity order, resulting in better outage performance.

\begin{figure}[pos=!t]
    \centering
    \includegraphics[width=5in]{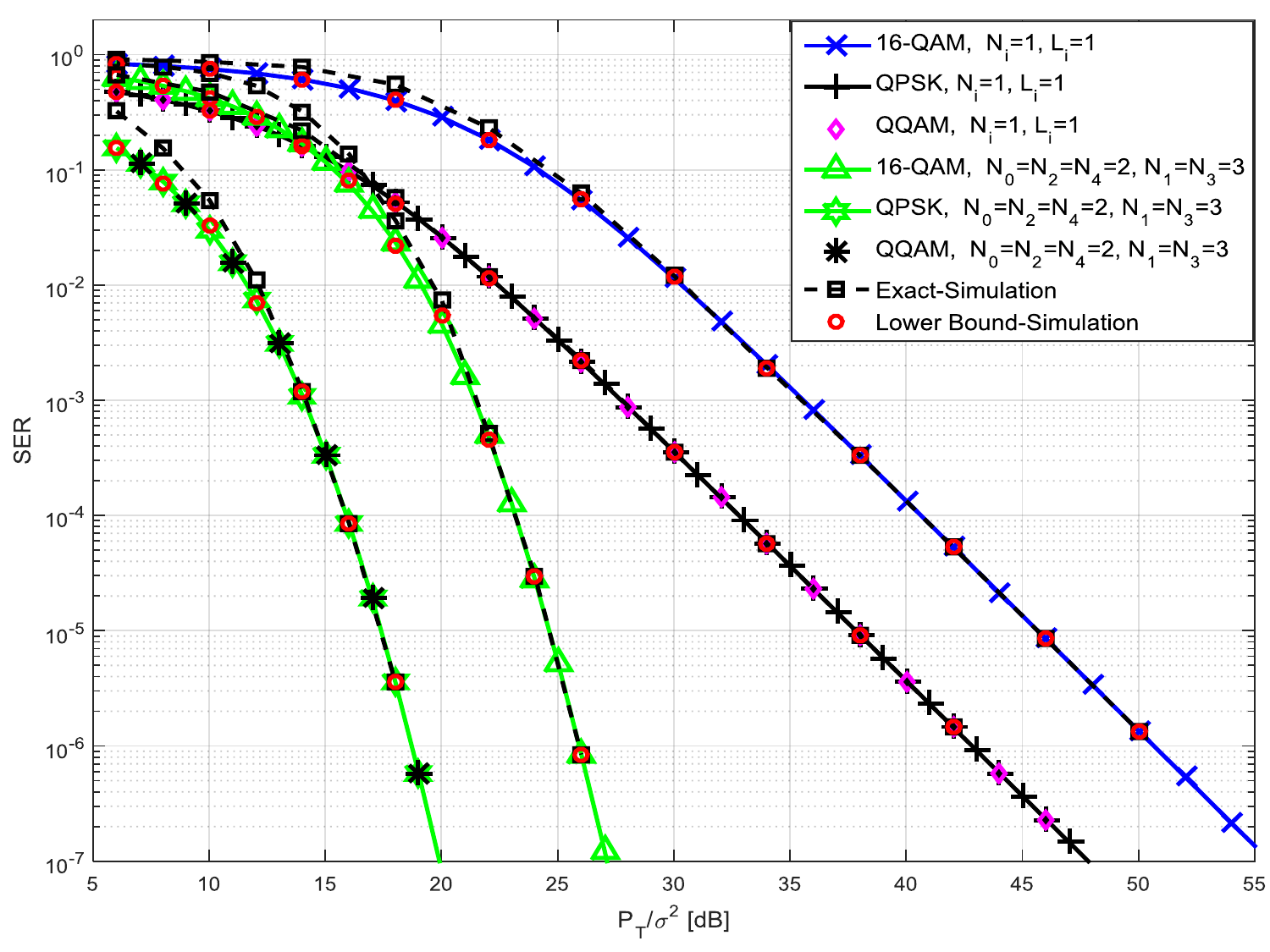}
    \caption{Multihop SER for different modulation schemes and antenna configurations.}
    \label{CMultiHopSER}
\end{figure}

Fig. \ref{CMultiHopSER} validates the SER lower-bound expression given in (\ref{Denklem_25}) for $M$-PSK and $M$-QAM. Total number of hops is 4. All shape factors are equal, namely, $m_i=2$. For QPSK, QQAM and 16-QAM, the cases without diversity ($N_i=1$ and $L_i=1$) are compared with the multi-antenna configurations $N_0=N_2=N_4=2$, $N_1=N_3=3$, and $L_i=2$, i.e., TAS/GSC is implemented in the first and third hops and TAS/MRC is carried out in the second and fourth hops. The analytical SER lower-bound curves are in excellent agreement with the corresponding lower-bound simulation results. Although a gap is observed between the exact SER simulation results and the lower-bound curves in the low-SNR region, they approach each other and eventually overlap as the SNR increases. This behavior is expected since the lower-bound expression approximates the end-to-end error performance using the weakest-hop SNR. As expected, the QPSK and QQAM curves overlap, since they correspond to equivalent four-point constellations under the same symbol-energy normalization. At a SER value of $10^{-5}$, the gain in total SNR due to the diversity techniques is about 20.5 dB for both QQAM and 16-QAM. This considerable gain demonstrates the effectiveness of antenna selection and diversity combining in DF multihop networks.

\section{Conclusion} \label{ChapVIConclusion}

In this paper, the performance of TAS/GSC is re-examined over i.i.d. Nakagami-$m$ fading channels developing an alternative finite-sum representation of the output-SNR distribution. This representation yields closed-form expressions for OP, SER, and ergodic capacity. To illustrate its applicability, the proposed analysis is extended to serial DF multihop networks over i.n.i.d. Nakagami-$m$ fading channels. Exact expressions for the end-to-end OP and ergodic capacity and an analytical lower bound on the average SER are obtained. Furthermore, high-SNR outage bounds are used to determine the diversity order. The analytical results show excellent agreement with Monte Carlo simulations, while the SER lower bound approaches the exact simulation results at high SNR. The numerical results reveal that the effect of the Nakagami-$m$ shape parameter on ergodic capacity depends on the antenna selection and combining scheme. For the considered GSC configurations, increasing $m$ slightly improves the ergodic capacity of MRC but decreases that of RAS, whereas the ergodic capacity of TAS/GSC decreases for the examined configurations. They also  demonstrate considerable OP and SER improvements through TAS/GSC and show that the diversity-limiting hop plays an important role in the end-to-end performance. 
These results demonstrate the applicability of the proposed analytical representation to the performance evaluation of TAS/GSC-assisted DF multihop networks. The developed finite-sum statistics may also provide a useful basis for investigating related multi-antenna relay configurations in emerging wireless networks, with appropriate extensions to the corresponding system and channel models.


\appendix

\makeatletter
\let\oldseccntformat\@seccntformat
\renewcommand{\@seccntformat}[1]{%
  \ifstrequal{#1}{section}%
    {Appendix~\thesection\@seccntDot\hskip 0.5em}%
    {\oldseccntformat{#1}}%
}
\makeatother
\section{Derivation of OP, SER, and Ergodic Capacity for GSC}  \label{ChapVISSROPDerivation}

In order to derive PDF and CDF of  $\beta_{k}^{i}$ (SNR of GSC), firstly its MGF is obtained and, in sequence, its PDF and CDF are derived from MGF. The MGF of $\beta_{k}^{i}$, $\Phi_{\beta_{k}^{i}}(s)$, can be obtained as \cite[eq. (5)]{ma2004efficient}
  \begin{equation} \label{Denklem_A2}
  \begin{split}
  \Phi_{\beta_{k}^{i}}(s)=L_i\binom{N_i}{L_i} \int_{0}^{\infty} e^{-sx}f_{\gamma_{j,k}^i}(x)[\Phi_{\gamma_{j,k}^i}(s,x)]^{L_i-1} [F_{\gamma_{j,k}^i}(x)]^{N_i-L_i}dx,
  \end{split}
  \end{equation}
where $\Phi_{\gamma_{j,k}^i}(s,x)$ is the complementary MGF of $\gamma_{j,k}^i$ \cite[eq. (3)]{ma2004efficient} and $F_{\gamma_{j,k}^i}(x)$ is the CDF of $\gamma_{j,k}^i$  given in (\ref{Denklem_3}). Inserting PDF given in (\ref{Denklem_2}) into (\ref{Denklem_A3}) and using the identity \cite[eq. (8.350.2)]{jeffrey2007table}, it turns to

  \begin{equation} \label{Denklem_A3}
  \begin{aligned}
  \Phi_{\gamma_{j,k}^i}(s,x)=&L_i\binom{N_i}{L_i} \int_{x}^{\infty}f_{\gamma_{j,k}^i}(y) e^{-sy}dy\\=&
  \frac{(m_i\lambda_i)^{m_i} \Gamma\left(m_i, (s+m_i\lambda_i)x\right) }{\Gamma(m_i) (s+m_i\lambda_i)^{m_i} }\\=&
  \frac{(m_i\lambda_i)^{m_i} e^{-(s+m_i\lambda_i)x} }{(s+m_i\lambda_i)^{m_i}}   \sum_{j=0}^{m_i-1} \left( \frac{ ((s+m_i\lambda_i)x)^j}{\Gamma(j+1)}\right).
  \end{aligned}
  \end{equation}
The equality in the last line is valid for integer values of $m_i$ \cite[eq. (8.352.2)]{jeffrey2007table}.

To proceed further, by using the identity provided in (\ref{Denklem_1}), one can simplify the terms $[\Phi_{\gamma_{j,k}^i}(s,x)]^{L_i-1}$ and  $[F_{\gamma_{j,k}^i}(x)]^{N_i-L_i}$ in (\ref{Denklem_A2}) as

  \begin{equation} \label{Denklem_A5}
  \begin{aligned}
 \left[\Phi_{\gamma_{j,k}^i}(s,x)\right]^{L_i-1}=&\left(\frac{(m_i\lambda_i)^{m_i} e^{-(s+m_i\lambda_i)x} }{(s+m_i\lambda_i)^{m_i}} \right)^{L_i-1} \left( \sum_{j=0}^{m_i-1} \left( \frac{ ((s+m_i\lambda_i)x)^j}{\Gamma(j+1)}\right) \right)^{L_i-1}\\=&
\left(\frac{(m_i\lambda_i)^{m_i} e^{-(s+m_i\lambda_i)x} }{(s+m_i\lambda_i)^{m_i}} \right)^{L_i-1} \sum_{q=0}^{(m_i-1)(L_i-1)} \left( \mu_i^{\Phi}(q, L_i-1, m_i-1) ((s+m_i\lambda_i)x)^q \right)
  \end{aligned}
  \end{equation}
    \begin{equation} \label{Denklem_A5_1}
  \begin{aligned}
      \left[F_{\gamma_{j,k}^i}(x)\right]^{N_i-L_i}=&\sum_{n=0}^{N_i-L_i} \binom{N_i-L_i}{n} (-1)^n e^{-nm_i\lambda_ix} \left( \sum_{j=0}^{m_i-1} \left( \frac{(m_i\lambda_ix)^j}{\Gamma(j+1)} \right)\right)^n \\
      =& \sum_{n=0}^{N_i-L_i} \sum_{j=0}^{n(m_i-1)} \left(\binom{N_i-L_i}{n} (-1)^n \mu_i^{F}\left(j, n, m_i-1\right)  x^j e^{-nm_i\lambda_ix} \right)
  \end{aligned}
  \end{equation}
where $a_{ik}^{\Phi}=\frac{u[m_i-1-k]}{\Gamma(k+1)}$ for $ \mu_i^{\Phi}(q, L_i-1, m_i-1)$ and $a_{ik}^{F}=\frac{(m_i\lambda_i)^k u[m_i-1-k]}{\Gamma(k+1)}$ for $ \mu_i^{F}(j, n, m_i-1)$. Note that polynomial expansion of $[F_{\gamma_{j,k}^i}(x)]^{N_i-L_i}$ is carried out after a binomial expansion. 
After all, substituting equivalences of $f_{\gamma_{j,k}^i}(x)$,  $[\Phi_{\gamma_{j,k}^i}(s,x)]^{L_i-1}$, and  $[F_{\gamma_{j,k}^i}(x)]^{N_i-L_i}$ into (\ref{Denklem_A2}) and making proper simplifications with the fact that the emerging integral is in the form of Gamma function,  $\Phi_{\beta_{k}^{i}}(s)$ turns to 
  \begin{equation} \label{Denklem_A6}
  \begin{aligned}
 \Phi_{\beta_{k}^{i}}(s)=&\sum_{n=0}^{N_i-L_i}  \sum_{q=0}^{(m_i-1)(L_i-1)}   \sum_{j=0}^{n(m_i-1)} \left( \frac{C_{\beta_{k}^{i}}^{\Phi} }{ (s+m_i\lambda_i)^{m_i(L_i-1)-q} (s+\lambda_{in})^{q+j+m_i} }\right)\\=&
 \sum_{q=0}^{(m_i-1)(L_i-1)}\left( \frac{C_{0,\beta_{k}^{i}}^{\Phi} }{ (s+m_i\lambda_i)^{m_i L_i}}\right)+
  \sum_{n=1}^{N_i-L_i} \sum_{q=0}^{(m_i-1)(L_i-1)}   \sum_{j=0}^{n(m_i-1)} \left( \frac{C_{\beta_{k}^{i}}^{\Phi} }{ (s+m_i\lambda_i)^{m_i(L_i-1)-q} (s+\lambda_{in})^{q+j+m_i} }\right)
  \end{aligned}
  \end{equation}
where $C_{\beta_{k}^{i}}^{\Phi}=\frac{(-1)^n(m_i\lambda_i)^{L_im_i} \Gamma(q+j+m_i)}{\Gamma(m_i) L_i^{q+j+m_i-1} }\binom{N_i}{L_i}\binom{N_i-L_i}{n}\mu_i^{\Phi}(q, L_i-1, m_i-1) \mu_i^{F}\left(j, n, m_i-1\right)$ and $\lambda_{in}=\frac{(L_i+n)m_i\lambda_i}{L_i}$. The second equality in (\ref{Denklem_A6}) is because of the fact that $\lambda_{i0}=m_i\lambda_i$ and, since $j=n=0$ and $\mu_i^{F}\left(0, 0, m_i-1\right)=1$,  $C_{0,\beta_{k}^{i}}^{\Phi}=\frac{(m_i\lambda_i)^{L_im_i} \Gamma(q+m_i)}{\Gamma(m_i) L_i^{q+m_i-1} }\binom{N_i}{L_i}\mu_i^{\Phi}(q, L_i-1, m_i-1)$.
Applying partial fraction decomposition to the term $\frac{1}{(s+\lambda_{i0})^{m_i(L_i-1)-q} (s+\lambda_{in})^{q+j+m_i}}$ in (\ref{Denklem_A6}) results into  
  \begin{equation} \label{Denklem_A7}
  \begin{aligned}
 \Phi_{\beta_{k}^{i}}(s)=&\sum_{q=0}^{(m_i-1)(L_i-1)}\left( \frac{C_{0,\beta_{k}^{i}}^{\Phi} }{ (s+\lambda_{i0})^{m_i L_i}}\right)+
  \sum_{n=1}^{N_i-L_i} \sum_{q=0}^{(m_i-1)(L_i-1)}  \sum_{j=0}^{n(m_i-1)} \sum_{p=1}^{m_i(L_i-1)-q} \left( \frac{C_{\lambda_{i0}}C_{\beta_{k}^{i}}^{\Phi} }{ (s+\lambda_{i0})^p}\right)\\&+
 \sum_{n=1}^{N_i-L_i} \sum_{q=0}^{(m_i-1)(L_i-1)}   \sum_{j=0}^{n(m_i-1)} \sum_{p=1}^{q+j+m_i} \left( \frac{C_{\lambda_{in}}C_{\beta_{k}^{i}}^{\Phi} }{(s+\lambda_{in})^p }\right),
  \end{aligned}
  \end{equation}
where $C_{\lambda_{i0}}$ and $C_{\lambda_{in}}$ are given in (\ref{Denklem_9}) and (\ref{Denklem_10}), respectively.

Taking inverse Laplace transform of (\ref{Denklem_A7}) gives PDF of $\beta_{k}^{i}$ as  
  \begin{equation} \label{Denklem_A9}
  \begin{aligned}
 f_{\beta_{k}^{i}}(x)=&\sum_{q=0}^{(m_i-1)(L_i-1)}\left(C_{0,\beta_{k}^{i}}^{\Phi} \frac{x^{m_i L_i-1} e^{-\lambda_{i0}x} }{\Gamma(m_i L_i) } \right)+
 \sum_{n=1}^{N_i-L_i} \sum_{q=0}^{(m_i-1)(L_i-1)}   \sum_{j=0}^{n(m_i-1)} \sum_{p=1}^{m_i(L_i-1)-q} \left(C_{\lambda_{i0}}C_{\beta_{k}^{i}}^{\Phi}\frac{x^{p-1} e^{-\lambda_{i0}x} }{\Gamma(p)}  \right)\\&+
 \sum_{n=1}^{N_i-L_i} \sum_{q=0}^{(m_i-1)(L_i-1)}   \sum_{j=0}^{n(m_i-1)} \sum_{p=1}^{q+j+m_i} \left(C_{\lambda_{in}}C_{\beta_{k}^{i}}^{\Phi} \frac{ x^{p-1} e^{-\lambda_{in}x}}{\Gamma(p) }   \right).
  \end{aligned}
  \end{equation}
Since this PDF is in the form of weighted Gamma PDFs, apparently, CDF of $\beta_{k}^{i}$ becomes straightforward as  
  \begin{equation} \label{Denklem_A10}
  \begin{aligned}
 F_{\beta_{k}^{i}}(x)=&\sum_{q=0}^{(m_i-1)(L_i-1)}\left(C_{0,\beta_{k}^{i}}^{\Phi} \lambda_{i0}^{-m_i L_i} \frac{\gamma(m_i L_i,\lambda_{i0}x)}{\Gamma(m_i L_i)} \right)+
 \sum_{n=1}^{N_i-L_i} \sum_{q=0}^{(m_i-1)(L_i-1)}   \sum_{j=0}^{n(m_i-1)} \sum_{p=1}^{m_i(L_i-1)-q} \left(C_{\lambda_{i0}}C_{\beta_{k}^{i}}^{\Phi}  \lambda_{i0}^{-p}  \frac{\gamma(p,\lambda_{i0}x) }{\Gamma(p)}  \right)\\&+
 \sum_{n=1}^{N_i-L_i} \sum_{q=0}^{(m_i-1)(L_i-1)}   \sum_{j=0}^{n(m_i-1)} \sum_{p=1}^{q+j+m_i} \left(C_{\lambda_{in}}C_{\beta_{k}^{i}}^{\Phi}   \lambda_{in}^{-p}  \frac{\gamma(p,\lambda_{in}x) }{\Gamma(p)}   \right).
  \end{aligned}
  \end{equation}
Using the identity \cite[eq. (8.352.1)]{jeffrey2007table} and the fact that constant terms sum up to $1$, $ F_{\beta_{k}^{i}}(x)$ can be rearranged as given in (\ref{Denklem_4}).

Ergodic capacity of GSC is derived based on CDF method \cite[eq. (45)]{yang2013performance} by using (4) as

  \begin{equation} \label{Denklem_A14}
  \begin{aligned}
 R_{GSC}=&\frac{\log_2(e)}{2}\int_{0}^{\infty} \frac{1-F_{\beta_{k}^{i}}(x)}{x+1}dx \\=&
 \frac{\log_2(e)}{2}\sum_{n=0}^{N_i-L_i}\sum_{j=0}^{T_{in} }  \left(c_{nj}^i \int_{0}^{\infty} \frac{x^je^{-\lambda_{in}x}}{x+1}dx\right), 
   \end{aligned}
  \end{equation}
The integral  in (\ref{Denklem_A14}) is defined and solved in  \cite[eq. (39)]{fidan2017performance} as 
  \begin{equation} \label{Denklem_A15}
  \begin{aligned}
  I_{EC}\left(m;\lambda\right)=\int_{0}^{\infty}\dfrac{x^m e^{-\lambda x}}{x+1}dx =
e^{\lambda}\Gamma(m+1)\Gamma(-m,\lambda),
  \end{aligned}
  \end{equation}
which completes the derivation of ergodic capacity. 

To unify the analysis through this work, another form of the MGF for GSC is obtained based on its CDF as \cite[eq. (17.12.3)]{jeffrey2007table} 
  \begin{equation} \label{Denklem_A16}
  \begin{aligned}
 \Phi_{\beta_{k}^{i}}(s)=1-\sum_{n=0}^{N_i-L_i} \sum_{j=0}^{T_{in} }  \left(c_{nj}^i \Gamma(j+1)\frac{s}{ (s+\lambda_{in})^{j+1}}\right).
   \end{aligned}
  \end{equation}
Hence, average SER of $M$-PSK and square $M$-QAM can be evaluated based on MGF method by $\sum\limits_{q=1}^{Q} a_q \int_{0}^{\theta_q} \Phi_{\beta_{k}^{i}}\left(\frac{\lambda_{\rm mod} }{\sin^2(\theta)}\right)d\theta$ \cite[eq. (5.3)]{simon2005digital}  as given in (\ref{Denklem_A17})
  \begin{equation} \label{Denklem_A17}
  \begin{aligned}
 P_{GSC}=\sum_{q=1}^{Q}\left(a_q\theta_q\right)-\sum_{q=1}^{Q} \sum_{n=0}^{N_i-L_i} \sum_{j=0}^{T_{in} }  \left(a_q c_{nj}^i \lambda_{\rm mod} \Gamma(j+1) \int_{0}^{\theta_q} \frac{\left(\sin^2(\theta)\right)^j}{ (\lambda_{\rm mod}+\lambda_{in}\sin^2(\theta))^{j+1}}d\theta\right),
   \end{aligned}
  \end{equation}
where the modulation type dependent parameters are given in Table \ref{SERTable}. The integral in (\ref{Denklem_A17}) is defined and solved as
  \begin{equation} \label{Denklem_A18}
  \begin{aligned}
 I_{m}(\theta_q; \lambda_{\rm mod}; \lambda_{in})=&\int_{0}^{\theta_q} \frac{\left(\sin^2(\theta)\right)^m}{ (\lambda_{\rm mod}+\lambda_{in}\sin^2(\theta))^{m+1}}d\theta\\=&
 \frac{1}{(2 \lambda_{\rm mod}+\lambda_{in})^{m+1}} \int_{0}^{2\theta_q} \frac{[1-\cos(\varepsilon)]^m}{[1-d\cos(\varepsilon)]^{m+1}}d\varepsilon\\=&
 \frac{1}{(\lambda_{\rm mod})^{1/2}(\lambda_{\rm mod}+\lambda_{in})^{m+1/2}} \int_{0}^{T} \sin^{2m}(\theta)d\theta
   \end{aligned}
  \end{equation}
where $T=\frac{1}{2}\tan^{-1}{\left(\frac{N}{D}\right)}+\frac{1}{\pi}\left(1-\sign(N)\frac{1+\sign(D)}{2}\right)$, $N=2\sqrt{\lambda_{\rm mod}(\lambda_{\rm mod}+\lambda_{in})}\sin(2\theta_q)$,  $D=(2\lambda_{\rm mod}+\lambda_{in})\cos(2\theta_q)-\lambda_{in}$ \cite[{eq. (5A.31) and (5A.32)}]{simon2005digital}, and $d=\lambda_{in}/ (2\lambda_{\rm mod}+\lambda_{in})$. The transformation from the first line to the second line in (\ref{Denklem_A18}) is accomplished  by using the identity $\sin^2(\theta)=[1-\cos(2\theta)]/2$ and, then,  employing change of variables $\varepsilon=2\theta$. The third line in (\ref{Denklem_A18}) is attained after two changes of variables: Firstly, Euler-Legendre change of variables, namely, $\cos(\varepsilon)=(d+\cos(x))/(1+d \cos(x))$ is implemented and, secondly, $x=2\theta$ is employed. Finally, the closed-form of the resulting integral in the last line of (\ref{Denklem_A18}) is defined and solved in  \cite[{eq. (38)}]{fidan2017performance} as
\begin{equation} \label{Denklem_A19}
\begin{aligned}
I_{m}(\varphi)=\int_{0}^{\varphi} \sin^{2m}(\theta)d\theta  =\dfrac{\sqrt{\pi}\Gamma[m+1/2]}{2\Gamma[m+1]}-\cos(\varphi)_2F_1(\dfrac{1}{2},\dfrac{1}{2}-m;\dfrac{3}{2};\cos^2(\varphi)).
\end{aligned}
\end{equation}
This integral is valid for $m>-1/2$ and $\sin(\varphi)\geq 0$ and $_2F_1(\iota,\tau;w;z)$ in (\ref{Denklem_A19}) is the Gauss hypergeometric function \cite[{eq. (9.111)}]{jeffrey2007table}.  For integer values of $m$, another closed-form is also available \cite[eq. (2.513.1)]{jeffrey2007table}.  So $I_{m}(\theta_q; \lambda_{\rm mod}; \lambda_{in})$ turns to (\ref{Denklem_13}).



\bibliographystyle{elsarticle-num}
\bibliography{references}

\end{document}